\documentclass[11pt,a4wide]{article}

\usepackage{amsmath,amssymb,amsfonts,epsf,psfrag,graphicx,array,verbatim,url,a4wide}
\usepackage[small,bf]{caption2}
\usepackage{multicol}
\usepackage[dvipsnames]{color}

\newtheorem{theorem}{Theorem}

\newtheorem{lemma}{Lemma}
\newtheorem{example}{Example}

\newenvironment{proof}{\noindent\emph{Proof.}\
}{\hfill\rule{8pt}{8pt}}
\newcommand{\PBS}[1]{\let\temp=\\#1\let\\=\temp}

\newcommand{\indicator}[1]{\mathbf{1}\left\{#1\right\}}

\begin{document}

\title{Optimal Scheduling of Peer-to-Peer File Dissemination}
\author{
Jochen Mundinger\footnote{EPFL-IC-LCA, BC203, Station 14, CH-1015 Lausanne,
Switzerland, Email: jochen.mundinger@epfl.ch}, \hspace{1mm} Richard
Weber\footnote{Statistical Laboratory, Centre for Mathematical Sciences,
Wilberforce Road, Cambridge CB3 0WB, UK} \hspace{1mm} and Gideon
Weiss\footnote{Department of Statistics, University of Haifa, Mount Carmel
31905, Israel}}

\date{}
\maketitle

\begin{abstract}
Peer-to-peer (P2P) overlay networks such as BitTorrent and Avalanche
are increasingly used for disseminating potentially large files from a
server to many end users via the Internet. The key idea is to divide
the file into many equally-sized parts and then let users download
each part (or, for network coding based systems such as Avalanche,
linear combinations of the parts) either from the server or from
another user who has already downloaded it.  However, their
performance evaluation has typically been limited to comparing one
system relative to another and typically been realized by means of
simulation and measurements. In contrast, we provide an analytic
performance analysis that is based on a new uplink-sharing version of
the well-known broadcasting problem.  Assuming equal upload
capacities, we show that the minimal time to disseminate the file is
the same as for the simultaneous send/receive version of the
broadcasting problem. For general upload capacities, we provide a
mixed integer linear program (MILP) solution and a complementary fluid
limit solution.  We thus provide a lower bound which can be used as a
performance benchmark for any P2P file dissemination system. We also
investigate the performance of a decentralized strategy, providing
evidence that the performance of necessarily decentralized P2P file
dissemination systems should be close to this bound and therefore that
it is useful in practice.
\end{abstract}

\section{Introduction}\label{s:efdintro}

Suppose that $M$ messages of equal length are initially known only at a single
source node in a network. The so-called \emph{broadcasting problem} is about
disseminating these $M$ messages to a population of $N$ other nodes in the
least possible time, subject to capacity constraints along the links of the
network. The assumption is that once a node has received one of the messages it
can participate subsequently in sending that message to its neighbouring nodes.

\subsection{P2P file dissemination background and related work}\label{s:efdintromot}

In recent years, overlay networks have proven a popular way of disseminating
potentially large files (such as a new software product or a video) from a
single server $S$ to a potentially large group of $N$ end users via the
Internet. A number of algorithms and protocols have been suggested, implemented
and studied. In particular, much attention has been given to peer-to-peer (P2P)
systems such as BitTorrent~\cite{c03}, Slurpie~\cite{sbb04},
SplitStream~\cite{cdknrs03}, Bullet'~\cite{krav03} and Avalanche~\cite{gr05},
to name but a few. The key idea is that the file is divided into $M$ parts of
equal size and that a given user may download any one of these (or, for network
coding based systems such as Avalanche, linear combinations of these) either
from the server or from a peer who has previously downloaded it. That is, the
end users collaborate by forming a P2P network of peers, so they can download
from one another as well as from the server. Our motivation for revisiting the
broadcasting problem is the performance analysis of such systems.\medskip

With the BitTorrent
protocol\footnote{http://bitconjurer.org/BitTorrent/protocol.html},
for example, when the load on the server is heavy, the protocol
delegates most of the uploading burden to the users who have already
downloaded parts of the file, and who can start uploading those parts
to their peers. File parts are typically $1/4$ megabyte (MB) in
size. An application helps downloading peers to find each other by
supplying lists of contact information about randomly selected peers
also downloading the file. Peers use this information to connect to a
number of neighbours. A full description can be found in
\cite{c03}. The BitTorrent protocol has been implemented successfully
and is deployed widely. A detailed measurement study of the BitTorrent
system is reported in \cite{pges05}. According to \cite{p04},
BitTorrent's share of the total P2P traffic has reached 53\% in June
2004. For recent measurements of the total P2P traffic on Internet
backbones see \cite{kbbcf04}.

Slurpie \cite{sbb04} is a very similar protocol, although, unlike BitTorrent,
it does not fix the number of neighbours and it adapts to varying bandwidth
conditions. Other P2P overlay networks have also been proposed. For example see
SplitStream \cite{cdknrs03} and Bullet' \cite{krav03}.

More recently,
Avalanche\footnote{http://www.research.microsoft.com/$\sim$pablo/avalanche.aspx}
\cite{gr05}, a scheme based on network coding \cite{acly00} has been
suggested.  Here, users download linear combinations of file parts
rather than individual file parts. This ensures that users do not need
to find specific parts in the system, but that any upload by a given
user can be of interest to any peer.  Thus, network coding can improve
performance in a decentralized scenario. Our results apply to any P2P
file dissemination system, whether or not it uses network coding.
\medskip

Performance analysis of P2P systems for file dissemination has typically been
limited to comparing one system relative to another and typically been realized
by means of simulation and measurements. We give the makespan, that is the
minimal time to fully disseminate the file, of $M$ parts from a server to $N$
end users in a centralized scenario. We thereby provide a lower bound which can
be used as a performance benchmark for any P2P file dissemination system. We
also investigate the part of the loss in efficiency that is due to the lack of
centralized control. Using a theoretical analysis, simulation as well as direct
computation, we show that even a naive randomized strategy disseminates the
file in an expected time that grows with $N$ in a similar manner to the minimal
time achieved with a centralized controller. This suggests that the performance
of necessarily decentralized P2P file dissemination systems should still be
close to our performance bound so that it is useful in practice.\medskip

In this paper, we provide the scheduling background, proofs and discussion of
the results in our extended abstracts \cite{mww05:perf} and \cite{mww06:mama}.
It is essentially Chapter 2 of \cite{m:phd}, but we have added Theorem
\ref{thm:decbound} and the part on theoretical bounds in Section
\ref{ss:efddecentreq}.
In~\cite{yd04} the authors also consider problems concerned with the service
capacity of P2P networks, however, they only give a heuristic argument for the
makespan with equal upload capacities when $N$ is of the simple form $2^n-1$.
In~\cite{qs04} a fluid model for BitTorrent-like networks is introduced and
studied, also looking at the effect of incentive mechanisms to address
free-riding. Link utilization and fairness are issues in~\cite{bhp05}.
In~\cite{mv04}, also motivated by the BitTorrent protocol and file swarming
systems in general, the authors consider a probabilistic model of coupon
replication systems. Multi-torrent systems are discussed in~\cite{gcxtdz05}.
There is other related work in~\cite{rs05}.

\subsection{Scheduling background and related work}\label{s:efdintrosched}

The broadcasting problem has been considered for different network topologies.
Comprehensive surveys can be found in \cite{hhl88} and \cite{hkmp95}.
On a complete graph, the problem was first solved in \cite{ct80} and
\cite{farley80}. Their communication model was a unidirectional telephone model
in which each node can either send or receive one message during each round,
but cannot do both. In this model, the minimal number of rounds required is
$2M-1+\lfloor\log_2{(N+1)}\rfloor$ for even $N$, and
$2M+\lfloor\log_2{(N+1)}\rfloor-\lfloor\frac{M-1+2^{\lfloor\log_2{(N+1)}\rfloor}}{(N+1)/2}\rfloor$
for odd $N$.\footnote{Bar-Noy, Kipnis and Schieber report a slightly different
expression in \cite{bks00}. This appears to be a transcription error in quoting
the result of Cockayne and Thomason.}

In \cite{bks00}, the authors considered the bidirectional telephone model in
which nodes can both send one message and receive one message simultaneously,
but they must be matched pairwise. That is, in each given round, a node can
only receive a message from the same node to which it sends a message. They
provide an optimal algorithm for odd $N$, which takes
$M+\lfloor\log_2{N}\rfloor$ rounds. For even $N$ their algorithm is optimal up
to an additive term of $3$, taking $M+\lfloor\log_2{N}\rfloor + M/N +2$ rounds.

The simultaneous send/receive model \cite{kc95} supposes that during each round
every user may receive one message and send one message. Unlike the telephone
model, it is not required that a user can send a message only to the same user
from which it receives a message. The optimal number of rounds turns out to be
$M+\lfloor\log_2{N}\rfloor$ and we will return to this result in Section
\ref{s:efdcentreq}.\medskip

In this paper, we are working with our new uplink-sharing model designed for
P2P file dissemination (cf. Section~\ref{s:efdmodel}). It is closely related to
the simultaneous send/receive model, but is set in continuous time. Moreover,
we permit users to have different upload capacities which are the constraints
on the data that can be sent per unit of time. This contrasts with previous
work in which the aim was to model interactions of processors and so it was
natural to assume that all nodes have equal capacities. Our work also differs
from previous work in that we are motivated by the evaluation of necessarily
decentralized P2P file dissemination algorithms, i.e., ones that can be
implemented by the users themselves, rather than by a centralized controller.
Our interest in the centralized case is as a basis for comparison and to give a
lower bound.
We show that in the case of equal upload capacities the optimal number of
rounds is $M+\lfloor\log_2{N}\rfloor$ as for the simultaneous send/receive
model. Moreover, we provide two complementary solutions for the case of general
upload capacities and investigate the performance of a decentralized strategy.

\subsection{Outlook}\label{s:efdintrooutlook}

The rest of this paper is organized as follows. In Section~\ref{s:efdmodel} we
introduce the uplink-sharing model and relate it to the simultaneous
send/receive model. Our optimal algorithm for the simultaneous send/receive
broadcasting problem is presented in Section~\ref{s:efdcentreq}.
We show that it also solves the problem for the uplink-sharing model with equal
capacities.
In Section~\ref{s:efdcentrgen} we show that the general uplink-sharing model
can be solved via a finite number of mixed integer linear programming (MILP)
problems. This approach is suitable for a small number of file parts $M$. We
provide additional insight through the solution of some special cases.
We then consider the limiting case that the file can be divided into infinitely
many parts and provide the centralized fluid solution. We extend these results
to the even more general situation where different users might have different
(disjoint) files of different sizes to disseminate
(Section~\ref{s:efdcentrfluidgen}). This approach is suitable for typical and
for large numbers of file parts $M$.
Finally, we turn to decentralized algorithms. In Section~\ref{ss:efddecentreq}
we evaluate the performance of a very simple and natural randomized strategy,
theoretically, by simulation and by direct computation. We provide results in
two different information scenarios with equal capacities showing that even
this naive algorithm disseminates the file in an expected time whose growth
rate with $N$ is similar to the growth rate of the minimal time that we have
found for a centralized controller. This suggests that the performance of
necessarily decentralized P2P file dissemination systems should still be close
to the performance bounds of the previous sections so that they are useful in
practice.
We conclude and present ideas for further research in Section~\ref{s:efddisc}.

\section{The Uplink-Sharing Model}\label{s:efdmodel}

We now introduce an abstract model for the file dissemination scenario
described in the previous section, focusing on the important features of P2P
file dissemination.\medskip

Underlying the file dissemination system is the Internet. Thus, each user can
connect to every other user and the network topology is a complete graph. The
server $S$ has upload capacity $C_S$ and the $N$ peers have upload capacities
$C_1,\ldots,C_N$, measured in megabytes per second (MBps). Once a user has
received a file part it can participate subsequently in uploading it to its
peers (source availability). We suppose that, in principle, any number of users
can simultaneously connect to the server or another peer, the available upload
capacity being shared equally amongst the open connections (fair sharing).
Taking the file size to be 1 MB, this means that if $n$ users try
simultaneously to download a part of the file (of size $1/M$) from the server
then it takes $n/MC_S$ seconds for these downloads to complete. Observe that
the rate at which an upload takes place can both increase and decrease during
the time of that upload (varying according to the number of other uploads with
which it shares the upload capacity), but we assume that uploads are not
interrupted until complete, that is the rate is always positive (continuity).
In fact, Lemma \ref{lem:singleupload} below shows that the makespan is not
increased if we restrict the server and all peers to carry out only a single
upload at a time.
We permit a user to download more than one file part simultaneously, but these
must be from different sources; only one file part may be transferred from one
user to another at the same time.  We ignore more complicated interactions and
suppose that the upload capacities, $C_S,C_1,\ldots,C_N$, impose the only
constraints on the rates at which file parts can be transferred between peers
which is a reasonable assumption if the underlying network is not overloaded.
Finally, we assume that rates of uploads and downloads do not constrain one
another.

Note that we have assumed the download rates to be unconstrained and this might
be considered unrealistic. However, we shall show a posteriori in
Section~\ref{s:efdcentreq} that if the upload capacities are equal then
additional download capacity constraints do not increase the minimum possible
makespan, as long as these download capacities are at least as big. Indeed,
this is usually the case in practice.

Typically, $N$ is the order of several thousands and the file size is up to a
few gigabytes (GB), so that there are several thousand file parts of size $1/4$
MB each.\medskip

Finding the minimal makespan looks potentially very hard as upload times are
interdependent and might start at arbitrary points in time. However, the
following two observations help simplify it dramatically. As we see in the next
section, they also relate the uplink-sharing model to the simultaneous
send/receive broadcasting model.

\begin{lemma}\label{lem:singleupload}
In the uplink-sharing model the minimal makespan is not increased by
restricting attention to schedules in which the server and each of the peers
only carry out a single upload at a time.
\end{lemma}

\begin{proof}
Identify the server as peer $0$ and, for each $i=0,1,\ldots,N$ consider the
schedule of peer $i$. We shall use the term \emph{job} to mean the uploading of
a particular file part to a particular peer. Consider the set of jobs, say $J$,
whose processing involves some sharing of the upload capacity $C_i$. Pick any
job, say $j$, in $J$ which is last in $J$ to finish and call the time at which
it finishes $t_f$.
Now fair sharing and continuity imply that job $j$ is amongst the last to start
amongst all the jobs finishing before or at time $t_f$. To see this, note that
if some job $k$ were to start later than $j$, then (by fair sharing and
continuity) $k$ must receive less processing than job $j$ by time $t_f$ and so
cannot have finished by time $t_f$. Let $t_s$ denote the starting time of job
$j$.

We now modify the schedule between time $t_s$ and $t_f$ as follows. Let $K$ be
the set of jobs with which job $j$'s processing has involved some sharing of
the upload capacity. Let us re-schedule job $j$ so that it is processed on its
own between times $t_f-1/C_iM$ and $t_f$. This consumes some amount of upload
capacity that had been devoted to jobs in $K$ between $t_f-1/C_iM$ and $t_f$.
However, it releases an exactly equal amount of upload capacity between times
$t_s$ and $t_f-1/C_iM$ which had been used by job $j$. This can now be
allocated (using fair sharing) to processing jobs in $K$.

The result is that $j$ can be removed from the set $J$. All jobs finish no
later than they did under the original schedule. Moreover, job $j$ starts later
than it did under the original schedule and the scheduling before time $t_s$
and after time $t_f$ is not affected. Thus, all jobs start no earlier than they
did under the original schedule. This ensures that the source availability
constraints are satisfied and that we can consider the upload schedules
independently. We repeatedly apply this argument until set $J$ is empty.
\end{proof}
\medskip

Using Lemma 1, a similar argument shows the following result.

\begin{lemma}\label{lem:idletime}
In the uplink-sharing model the minimal makespan is not increased by
restricting attention to schedules in which uploads start only at times that
other uploads finish or at time $0$.
\end{lemma}

\begin{proof}
By the previous Lemma it suffices to consider schedules in which the server and
each of the peers only carry out a single upload at a time. Consider the joint
schedule of all peers $i=0,1,\ldots,N$ and let $J$ be the set of jobs that
start at a time other than $0$ at which no other upload finishes. Pick a job,
say $j$, that is amongst the first in $J$ to start, say at time $t_s$. Consider
the greatest time $t_f$ such that $t_f<t_s$ and $t_f$ is either $0$ or the time
that some other upload finishes and modify the schedule so that job $j$ already
starts at time $t_f$.

The source availability constraints are still satisfied and all uploads finish
no later than they did under the original schedule. Job $j$ can be removed from
the set $J$ and the number of jobs in $J$ that start at time $t_s$ is decreased
by 1, although there might now be more (but at most $N$ in total) jobs in $J$
that start at the time that job $j$ finished in the original schedule. But this
time is later than $t_s$. Thus, we repeatedly apply this argument until the
number of jobs in $J$ that start at time $t_s$ becomes $0$ and then move along
to jobs in $J$ that are now amongst the first in $j$ to start at time
$t_s'>t_s$. Note that once a job has been removed from $J$, it will never be
included again. Thus we continue until the set $J$ is empty.
\end{proof}

\section{Centralized Solution for Equal Capacities}\label{s:efdcentreq}

In this section, we give the optimal centralized solution of the uplink-sharing
model of the previous section with equal upload capacities. We first consider
the simultaneous send/receive broadcasting model in which the server and all
users have upload capacity of 1. The following theorem provides a formula for
the minimal makespan and a centralized algorithm that achieves it is contained
in the proof.

This agrees with a result of Bar-Noy, Kipnis and Schieber~\cite{bks00}, who
obtained it as a by-product of their result on the bidirectional telephone
model. However, they required pairwise matchings in order to apply the results
from the telephone model. So, for the simultaneous send/receive model, too,
they use perfect matching in each round for odd $N$, and perfect matching on
$N-2$ nodes for even $N$. As a result, their algorithm differs for odd and even
$N$ and it is substantially more complicated, to describe, implement and prove
to be correct, than the one we present within the proof of
Theorem~\ref{thm:makespan}.
Theorem~\ref{thm:makespan} has been obtained also by Kwon and Chwa~\cite{kc95},
via an algorithm for broadcasting in hypercubes. By contrast, our explicitly
constructive proof makes the structure of the algorithm very easy to see.
Moreover, it makes the proof of Theorem~\ref{thm:makespanuplink}, that is, the
result for the uplink-sharing model, a trivial consequence (using Lemmata
\ref{lem:singleupload} and \ref{lem:idletime}).

Essentially, the $\log_2{N}$-scaling is due to the P2P approach. This compares
favourably to the linear scaling of $N$ that we would obtain for a fixed set of
servers. The factor of $1/M$ is due to splitting the file into parts.

\begin{theorem}
In the simultaneous send/receive model with all upload and download capacities
equal to 1, the minimum number of rounds is $M+\lfloor\log_2{N}\rfloor$, each
round taking up $1/M$ units of time. Equivalently, for all $M$, $N$, the
minimal makespan is
\begin{equation}
T^*=1+\frac{\lfloor\log_2{N}\rfloor}{M}\,.\label{e:makespan}
\end{equation}
\label{thm:makespan}
\end{theorem}

\begin{proof}
Suppose that $N=2^n-1+x$, for $x=1,\ldots,2^n$. So $n=\lfloor\log_2{N}\rfloor$.
The fact that $M+n$ is a lower bound on the number of rounds is
straightforwardly seen as follows. There are $M$ different file parts and the
server can only upload one file part (or one linear combination of file parts)
in each round. Thus, it takes at least $M$ rounds until the server has made
sufficiently many uploads of file parts (or linear combinations of file parts)
that the whole file can be recovered. The last of these $M$ uploads by the
server contains information that is essential to recovering the file, but this
information is now known to only the server and one peer. It must takes at
least $n$ further rounds to disseminate this information to the other $N-1$
peers.\medskip

Now we show how the bound can be achieved. The result is trivial for $M=1$. It
is instructive to consider the case $M=2$ explicitly.  If $n=0$ then $N=1$ and
the result is trivial.  If $n=1$ then $N$ is $2$ or $3$. Suppose $N=3$. In the
following diagram each line corresponds to a round; each column to a peer. The
entries denote the file part that the peer downloads that round. The bold
entries indicate downloads from the server; un-bold entries indicate downloads
from a peer who has the corresponding part.

{\footnotesize \setlength{\arraycolsep}{0.2mm}
\renewcommand{\arraystretch}{0.9}
\[
\begin{array}{*{28}{>{\PBS\centering$}p{.45cm}<{$}}}
{\bf 1}\\
&{\bf 2}&1\\
2&1&{\bf 2}\\
\end{array}
\]}

\noindent Thus, dissemination of the two file parts to the 3 users can be
completed in 3 rounds. The case $N=2$ is even easier.

If $n \geq 2$, then in rounds $2$ to $n$ each user uploads his part to a peer
who has no file part and the server uploads part $2$ to a peer who has no file
part. We reach a point, shown below, at which a set of $2^{n-1}$ peers have
file part 1, a set of $2^{n-1}-1$ peers have file part 2, and a set of $x$
peers have no file part (those denoted by $*\cdots*$). Let us call these three
sets $A_1$, $A_2$ and $A_0$, respectively.\medskip

{\footnotesize \setlength{\arraycolsep}{0.2mm}
\renewcommand{\arraystretch}{0.9}
\[
\begin{array}{*{28}{>{\PBS\centering$}p{.45cm}<{$}}}
{\bf 1}\\
&{\bf 2}&1\\
&&&{\bf 2}&1&2&1\\
&&&&&&&{\bf 2}&1&2&1&2&1&2&1\\
&&&&&&&&&&&&&&&\vdots\\
&&&&&&&&&&&&&&&&&&&&&&{\bf 2}&1&\cdots&2&1&\multicolumn{1}{c}{*\cdots*}\\
\end{array}
\]}

\medskip\noindent In round $n+1$ we let peers in $A_1$ upload part 1 to
$2^{n-1}-\lfloor x/2 \rfloor$  peers in $A_2$ and to $\lfloor x/2 \rfloor$
peers in $A_0$ (If $x=1$, to $2^{n-1}-1$ peers in $A_2$ and to $1$ peer in
$A_0$). Peers in $A_2$ upload part 2 to $2^{n-1}-\lceil x/2 \rceil$ peers in
$A_1$ and to another $\lceil x/2 \rceil -1$ peers in $A_0$. The server uploads
part 2 to a member of $A_0$ (If $x=1$, to a member of $A_1$). Thus, at the end
of this round $2^n-x$ peers have both file parts, $x$ peers have only file part
1, and $x-1$ peers have only file part 2. One more round (round $n+2$) is
clearly sufficient to complete the dissemination.\medskip \smallskip

Now consider $M\geq 3$. The server uploads part 1 to one peer in round 1. In
rounds $j=2,\ldots,\min\{n,M-1\}$, each peer who has a file part uploads his
part to another peer who has no file part and the server uploads part $j$ to a
peer who has no file part.  If $M\leq n$, then in rounds $M$ to $n$ each peer
uploads his part to a peer who has no file part and the server uploads part $M$
to a peer who has no file part. As above, we illustrate this with a diagram.
Here we show the first $n$ rounds in the case $M\leq n$.

{\footnotesize \setlength{\arraycolsep}{0.2mm}
\renewcommand{\arraystretch}{0.9}
\[
\begin{array}{*{28}{>{\PBS\centering$}p{.45cm}<{$}}}
{\bf 1}\\ &{\bf 2}&1\\ &&&{\bf 3}&1&2&1\\ &&&&&&&{\bf
4}&1&2&1&3&1&2&1\\ &&&&&&&&&&&&&&&\vdots\\ &&&&&&&&&&&&&&&&{\bf
M}&1&\cdots&2&1\\ &&&&&&&&&&&&&&&&&&&&&\vdots\\
&&&&&&&&&&&&&&&&&&&&&&{\bf
M}&1&\cdots&2&1&\multicolumn{1}{c}{*\cdots*}\\
\end{array}
\]}

\noindent When round $n$ ends, $2^n-1$ peers have one file part and $x$ peers
have no file part. The number of peers having file part $i$ is given in the
second column of Table~\ref{t:filepartnumbers}. In this table any entry which
evaluates to less than 1 is to be read as $0$ (so, for example, the bottom two
entries in column 2 and the bottom entry in column 3 are 0 for $n=M-2$). Now in
round $n+1$, by downloading from every peer who has a file part, and
downloading part $\min\{n+1,M\}$ from the server, we can obtain the numbers
shown in the third column. Moreover, we can easily arrange so that peers can be
divided into the sets $B_{12}$, $B_{1p}$, $B_1$, $B_2$ and $B_p$ as shown in
Table~\ref{t:filepartsets}.
\begin{table}[]\small
\renewcommand{\arraystretch}{1.2}
\[
\begin{array}{cllllcl}
\hline
\text{Part} & \multicolumn{6}{c}{\text{Numbers of the file parts at the ends of rounds}}\\
  & {n} & {n+1} & {n+2} & {n+3} & \cdots &{n+M-1}\\
\hline
1 & 2^{n-1} & 2^n & N & N &\cdots &N \\
2 & 2^{n-2} & 2^{n-1} & 2^n & N & \cdots &N \\
3 & 2^{n-3} & 2^{n-2} & 2^{n-1} & 2^n &\cdots &N \\
4 & 2^{n-4} & 2^{n-3} & 2^{n-2} & 2^{n-1} &\cdots &N \\
\vdots & \vdots & \vdots & \vdots & \vdots&&\vdots \\
M-2 & 2^{n-M+2} & 2^{n-M+3} & 2^{n-M+4} & 2^{n-M+5} &\cdots &N \\
M-1 & 2^{n-M+1} & 2^{n-M+2} & 2^{n-M+3} & 2^{n-M+4} & \cdots &2^{n} \\
M & 2^{n-M+1}-1 & 2^{n-M+2}-1 &2^{n-M+3}-1 & 2^{n-M+4}-1 &\cdots &2^{n}-1\\\hline
\end{array}
\]
\caption{Number of file part replica as obtained with our algorithm.}\label{t:filepartnumbers}
\end{table}
\begin{table}[]
\begin{center}
\begin{tabular}{lll}
\hline
set      &  peers in the set have                 & number of peers in set\\
\hline
$B_{12}$ &  parts 1 and 2                       & $2^{n-1}-\lfloor x/2\rfloor$\\
$B_{1p}$ &  part 1 and a part other than 1 or 2 & $2^{n-1}-\lceil x/2\rceil$\\
$B_{1}$  &  just part 1                         & $x$ \\
$B_{2}$  &  just part 2                         & $\lfloor x /2\rfloor$\\
$B_{p}$  &  just a part other than 1 or 2       & $\lceil x/2\rceil-1$\\\hline
\end{tabular}
\end{center}
\caption{File parts held by various sets of peers at the end of round $n+1$.}\label{t:filepartsets}
\end{table}
In round $n+2$, $x-1$ of the peers in $B_{1}$ upload part 1 to peers in $B_{2}$
and $B_{p}$. Peers in $B_{12}$ and $B_{2}$ each upload part 2 to the peers in
$B_{1p}$ and to $\lceil x/2\rceil$ of the peers in $B_{1}$. The server and the
peers in $B_{1p}$ and $B_{p}$ each upload a part other than 1 or 2 to the peers
in $B_{12}$ and to the other $\lfloor x/2\rfloor$ peers in $B_{1}$. The server
uploads part $\min\{n+2,M\}$ and so we obtain the numbers in the fourth column
of Table~\ref{t:filepartnumbers}. Now all peers have part 1 and so it can be
disregarded subsequently. Moreover, we can make the downloads from the server,
$B_{1p}$ and $B_{p}$ so that (disregarding part 1) the number of peers who
ultimately have only part 3 is $\lfloor x /2\rfloor$. This is possible because
the size of $B_{p}$ is no more than $\lfloor x /2\rfloor$; so if $j$ peers in
$B_{p}$ have part $3$ then we can upload part 3 to exactly $\lfloor x /2\rfloor-j$
peers in $B_{1}$.
Thus, a similar partitioning into sets as in Table~\ref{t:filepartsets} will
hold as we start step $n+3$ (when parts 2 and 3 takes over the roles of parts 1
and 2 respectively).

We continue similarly in subsequent rounds, until at the end of round $n+M-1$,
all peers have parts $1,\ldots,M-2$, $2^n-x$ peers also have both part $M-1$
and part $M$, $x$ peers also have only part $M-1$, and $x-1$ peers also have
only part $M$. It now takes just a final round to ensure that all peers have
parts $M-1$ and $M$.
\end{proof}
\smallskip

Note that the optimal strategy above follows two principles. As many different
peers as possible obtain file parts early on so that they can start uploading
themselves and the maximal possible upload capacity is used. Moreover, there is
a certain balance in the upload of different file parts so that no part gets
circulated too late.\medskip

It is interesting that not \emph{all} the available upload capacity is used.
Suppose $M\geq 2$. Observe that in round $k$, for each $k=n+2,\ldots, n+M-1$,
only $x-1$ of the $x$ peers (in set $B_{1}$) who have only file part $k-n-1$
make an upload. This happens $M-2$ times. Also, in round $n+M$ there are only
$2x-1$ uploads, whereas $N+1$ are possible. Overall, we use $N+M-2x$ less
uploads than we might. It can be checked that this number is the same for
$M=1$.

Suppose we were to follow a schedule that uses only $x$ uploads during
round $n+1$, when the last peer gets its first file part. We would be
using $2^n-x$ less uploads than we might in this round. Since
$2^n-x\leq N+M-2x$, we see that the schedule used in the proof above
wastes at least as many uploads. So the mathematically interesting
question arises as to whether or not it is necessary to use more than
$x$ uploads in round $n+1$. In fact, $(N+M-2x)-(2^n-x) = M-1$, so, in
terms of the total number of uploads, such a scheduling could still
afford not to use one upload during each of the last $M-1$ rounds.
The question is whether or not each file part can be made available
sufficiently often.

The following example shows that if we are not to use more than $x$ uploads in
round $n+1$ we will have to do something quite subtle. We cannot simply pick
any $x$ out of the $2^n$ uploads possible and still hope that an optimal
schedule will be \emph{shiftable}: by which we mean that the number of copies
of part $j$ at the end of round $k$ will be the same as the number of copies of
part $j-1$ at the end of round $k-1$. It is the fact that the optimal schedule
used in Theorem~\ref{thm:makespan} is shiftable that makes its optimality so
easy to see.

\begin{example}
Suppose $M=4$ and $N=13=2^3+6-1$, so $M+\lfloor \log_2 N\rfloor = 7$. If we
follow the same schedule as in Theorem~\ref{thm:makespan}, we reach after round
$3$,
\[
\setlength{\arraycolsep}{0.5mm} \renewcommand{\arraystretch}{0.5}
\begin{array}{*{25}{>{\PBS\centering$}p{.45cm}<{$}}}
{\bf 1}\\[-6pt] &{\bf 2}&1\\[-6pt] &&&{\bf 3}&1&2&1\\[-3pt]
&&&&&&&\cdot&\cdot&\cdot&\cdot&\cdot&\cdot\\[-6pt]
\end{array}\]
Now if we only make $x=6$ uploads during round $4$, then there are eight ways
to choose which six parts to upload and which two parts not to upload. One can
check that in no case is it possible to arrange so that once this is done and
uploads are made for round $5$ then the resulting state has the same numbers of
parts $2$, $3$ and $4$, respectively, as the numbers of parts $1$, $2$ and $3$
at the end of round $4$. That is, there is no shiftable optimal schedule. In
fact, if our six uploads has been four part $1$s and two part $2$s, then it
would not even be possible to achieve \eqref{e:makespan}.
\end{example}

In some cases, we can achieve \eqref{e:makespan}, if we relax the demand that
the schedule be shiftable. Indeed, we conjecture that this is always possible
for at least one schedule that uses only $x$ uploads during round $n+1$.
However, the fact that we cannot use essentially the same strategy in each
round makes the general description of a non-shiftable optimal schedule very
complicated. Our aim has been to find an optimal (shiftable) schedule that is
easy to describe. We have shown that this is possible if we do use the spare
capacity at round $n+1$. For practical purposes this is desirable anyway, since
even if it does not affect the makespan it is better if users obtain file parts
earlier.
\medskip

When $x=2^n$ our schedule can be realized using matchings between the $2^n$
peers holding the part that is to be completed next and the server together
with the $2^n-1$ peers holding the remaining parts. But otherwise this is not
always possible to schedule only with matchings.
This is why our solution would not work for the more constrained telephone-like
model considered in \cite{bks00} (where, in fact, the answer differs as $N$ is
even or odd).
to describe.
\medskip

The solution of the simultaneous send/receive broadcasting model problem now
gives the solution of our original uplink-sharing model when all capacities are
the same.

\begin{theorem} Consider the uplink-sharing model with all upload capacities equal to 1. The minimal makespan is given by \eqref{e:makespan}, for
all $M$, $N$, the same as in the simultaneous send/receive model with
all upload capacities equal to 1.
\end{theorem}

\begin{proof}
Note that under the assumptions of the theorem and with application of
Lemmas~\ref{lem:singleupload} and \ref{lem:idletime}, the optimal solution to
the uplink-sharing model is the same as that of the simultaneous send/receive
broadcast model when all upload capacities equal to 1.
\end{proof}
\medskip

In the proof of Theorem~\ref{thm:makespan} we explicitly gave an optimal
schedule which also satisfies the constraints that no peer downloads more than
a single file part at a time. Thus, we also have the following result.

\begin{theorem}
In the uplink-sharing model  with all upload capacities equal to 1,
constraining the peers' download rates to 1 does not further increase the
minimal makespan.
\label{thm:makespanuplink}
\end{theorem}

\section{Centralized Solution for General Capacities}\label{s:efdcentrgen}

We now consider the optimal centralized solution in the general case of the
uplink-sharing model in which the upload capacities may be different.
Essentially, we have an unusual type of precedence-constrained job scheduling
problem. In Section~\ref{ss:efdcentrgenlp} we formulate it as a mixed integer
linear program (MILP). The MILP can also be used to find approximate solutions
of bounded size of sub-optimality. In practice, it is suitable for a small
number of file parts $M$. We discuss its implementation in
Section~\ref{ss:efdcentrgenlpdisc}. Finally, we provide additional insight into
the solution with different capacities by considering special choices for $N$
and $M$ when $C_1=C_2=\cdots=C_N$, but $C_S$ might be different (Sections
\ref{ss:efdcentrgenspecial} and \ref{ss:efdcentrfluidgenspecial}).

\subsection{MILP formulation}\label{ss:efdcentrgenlp}

In order to give the MILP formulation, we need the following Lemma.
Essentially, it shows that time can be discretized suitably.

\begin{lemma}\label{lem:tau}
Consider the uplink-sharing model and suppose all uplink capacities are integer
multiples of a common time unit. Then there exists $\tau$, such that under an
optimal schedule all uploads start and finish at integer multiples of $\tau$.
\end{lemma}

\begin{proof}
Rescale time so that $C_S,C_1,\ldots,C_N$ are all integers and let $L$ be their
least common multiple. The time that the first job completes must be an integer
multiple of $1/ML$. All remaining jobs are of sizes $1/M$ or $1/M - (1/M
C_j)C_i$ for various $C_i\leq C_j$. These are also integer multiples of $1/ML$.
Repeating this, we find that the time that the second job completes, and the
lengths of all remaining jobs at this point must be integer multiples of
$1/(ML)^2$. Repeating further, we find that $\tau=1/(ML)^{MN}$ suffices.
\end{proof}\medskip

We next show how the solution to the general problem can be found by solving a
number of linear programs. Let time interval $t$ be the interval $[t\tau,
t\tau+\tau)$, $t=0,\ldots\ $. Identify the server as peer $0$. Let $x_{ijk}(t)$
be 1 or 0 as peer $i$ downloads file part $k$ from peer $j$ during interval $t$
or not. Let $p_{ik}(t)$ denote the proportion of file part $k$ that peer $i$
has downloaded by time $t$. Our problem is then is to find the minimal $T$ such
that the optimal value of the following MILP is $MN$. Since this $T$ is
certainly greater than $1/C_S$ and less than $N/C_S$, we can search for its
value by a simple bisection search, solving this LP for various $T$:
\begin{equation}
\text{maximize}\quad \sum_{i,k} p_{ik} (T)
\end{equation}
subject to the constraints given below. The source availability constraint
(\ref{e:source}) guarantees that a user has completely downloaded a part before
he can upload it to his peers. The connection constraint (\ref{e:connection})
requires that each user only carries out a single upload at a time. This is
justified by Lemma \ref{lem:singleupload} which also saves us another essential
constraint and variable to control the actual download rates: The single user
downloading from peer $j$ at time $t$ will do so at rate $C_j$ as expressed in
the link constraint (\ref{e:link}). Continuity and stopping constraints
(\ref{e:continuity}, \ref{e:stopping}) require that a download that has started
will not be interrupted until completion and then be stopped. The exclusivity
constraint (\ref{e:exclusivity}) ensures that each user downloads a given file
part only from one peer, not from several ones. Stopping and exclusivity
constraints are not based on assumptions, but obvious constraints to exclude
redundant uploads.\medskip

\noindent \textbf{Regional constraints}
\begin{gather}
x_{ijk}(t) \in \{0,1\} \text{ for all } i,j,k,t\\
p_{ik}(t) \in [0,1] \text{ for all } i,k,t
\end{gather}

\noindent \textbf{Link constraints between variables}
\begin{gather}\label{e:link}
p_{ik}(t) = M \tau \sum_{t'=0}^{t-\tau} \sum_{j=0}^N x_{ijk}(t')C_j\, \text{ for all } i,k,t
\end{gather}

\noindent \textbf{Essential constraints}
\begin{gather}
x_{ijk}(t)-\xi_{jk}(t) \leq 0\, \text{ for all } i,j,k,t \quad \text{(Source availability constraint)}\label{e:source}\\
\sum_{i,k} x_{ijk}(t) \leq 1\, \text{ for all } j,t \quad \text{(Connection constraint)}\label{e:connection}\\
x_{ijk}(t)-\xi_{ik}(t+1)-x_{ijk}(t+1) \leq 0\, \text{ for all } i,j,k,t \quad \text{(Continuity constraint)}\label{e:continuity}\\
x_{ijk}(t)+\xi_{ik}(t) \leq 1\, \text{ for all } i,j,k,t \quad \text{(Stopping constraint)}\label{e:stopping}\\
\sum_{j} x_{ijk}(t) \leq 1\, \text{ for all } i,k,t \quad \text{(Exclusivity constraint)}\label{e:exclusivity}
\end{gather}

\noindent \textbf{Initial conditions}
\begin{gather}
p_{0k}(0)=1 \text{ for all } k\\ p_{ik}(0)=0 \text{ for all } i,k
\end{gather}

\noindent Constraints \eqref{e:continuity}--\eqref{e:source} have been
linearized. Background can be found in \cite{b:s02}. For this, we used the
auxiliary variable $\xi_{ik}(t)=\indicator{p_{ik}(t) = 1}$. This definition can
be expressed through the following linear constraints.\medskip

\noindent \textbf{Linearization constraints}
\begin{gather}
\xi_{ik}(t) \in\{0,1\}\text{ for all } \, i,k,t\\
p_{ik}(t) - \xi_{ik}(t) \geq 0 \text{ and } p_{ik}(t) - \xi_{ik}(t) < 1\, \text{ for all } i,k,t
\end{gather}
It can be checked that together with \eqref{e:continuity}--\eqref{e:source},
indeed, this gives
\begin{gather}
x_{ijk}(t)=1 \text{ and } p_{ik}(t+1) < 1\Longrightarrow x_{ijk}(t+1)=1 \text{ for all } i,j,k,t\\
p_{ik}(t)=1 \Longrightarrow x_{ijk}(t)=0 \text{ for all }i,j,k,t\\
p_{jk}(t) < 1 \Longrightarrow x_{ijk}(t)=0 \text{ for all } i,j,k,t
\end{gather}
that is, continuity, stopping and source availability constraints respectively.

\subsection{Implementation of the MILP}\label{ss:efdcentrgenlpdisc}

MILPs are well-understood and there exist efficient computational methods and
program codes. The simplex method introduced by Dantzig in 1947, in particular,
has been found to yield an efficient algorithm in practice as well as providing
insight into the theory. Since then, the method has been specialized to take
advantage of the particular structure of certain classes of problems and
various interior point methods have been introduced.
For integer programming there are branch-and-bound, cutting plane
(branch-and-cut) and column generation (branch-and-price) methods as well as
dynamic programming algorithms. Moreover, there are various approximation
algorithms and heuristics.
These methods have been implemented in many commercial optimization libraries
such as OSL or CPLEX. For further reading on these issues the reader is
referred to \cite{b:nw88}, \cite{b:bt97} and \cite{b:z04}.

Thus, implementing and solving the MILPs gives the minimal makespan solution.
Although, as the numbers of variables and constraints in the LP grows
exponentially in $N$ and $M$, this approach is not practical for large $N$ and
$M$.

Even so, we can use the LP formulation to obtain a bounded approximation to the
solution. If we look at the problem with a greater $\tau$, then the job end and
start times are not guaranteed to lie at integer multiples of $\tau$. However,
if we imagine that each job does take until the end of an $\tau$-length
interval to finish (rather than finishing before the end), then we will
overestimate the time that each job takes by at most $\tau$. Since there are
$NM$ jobs in total, we overestimate the total time taken by at most $NM\tau$.
Thus, the approximation gives us an upper bound on the time taken and is at
most $NM\tau$ greater than the true optimum. So we obtain both upper and lower
bounds on the minimal makespan. Even for this approximation, the computing
required is formidable for large $N$ and $M$.

\subsection{Insight for special cases with small $N$ and $M$}\label{ss:efdcentrgenspecial}

We now provide some insight into the minimal makespan solution with different
capacities by considering special choices for $N$ and $M$ when
$C_1=C_2=\cdots=C_N$, but $C_S$ might be different. This addresses the case of
the server having a significantly higher upload capacity than the end
users.\medskip

Suppose $N=2$ and $M=1$, that is, the file has not been split. Only the server
has the file initially, thus either (a) both peers download from the server, in
which case the makespan is $T=2/C_S$, or (b) one peer downloads from the server
and then the second peer downloads from the first; in this case $T=1/C_S +
1/C_1$. Thus, the minimal makespan is $T^*={1}/{C_S} +
\min\{{1}/{C_S},{1}/{C_1}\}$.

If $N=M=2$ we can again adopt a brute force approach. There are 16 possible
cases, each specifying the download source that each peer uses for each part.
These can be reduced to four by symmetry.\medskip

\noindent {\bf Case A:} Everything is downloaded from the server. This is
effectively the same as case (a) above. When $C_1$ is small compared to $C_S$,
this is the optimal strategy.

\noindent {\bf Case B:} One peer downloads everything from the server. The
second peer downloads from the first. This is as case (b) above, but since the
file is split in two, $T$ is less.

\noindent {\bf Case C:} One peer downloads from the server. The other peer
downloads one part of the file from the server and the other part from the
first peer.

\noindent {\bf Case D:} Each peer downloads exactly one part from the server
and the other part from the other peer. When $C_1$ is large compared to $C_S$,
this is the optimal strategy.

In each case, we can find the optimal scheduling and hence the minimal
makespan. This is shown in Table~\ref{t:makespan}.

\begin{table}[h]
\[
\renewcommand{\arraystretch}{1.2}
\begin{array}{cll}\hline
\text{case} & \text{makespan} \\ \hline
\text{A} & \frac{2}{C_S} \\
\text{B} & \frac{1}{2C_S}+ \frac{1}{2C_1} + \max\left(\frac{1}{2C_S},\frac{1}{2C_1}\right) \\
\text{C} & \frac{1}{2C_S} + \max\left(\frac{1}{C_S},\frac{1}{2C_1}\right) \\
\text{D} & \frac{1}{C_S} + \frac{1}{2C_1} \\ \hline
\end{array}
\]
\caption{Minimal makespan in the four possible cases when $N=M=2$.}\label{t:makespan}
\end{table}

\begin{figure}[h]
\begin{center}
\begin{psfrags}
\psfrag{Time}{$\dfrac{\text{Time}}{C_s}$} \psfrag{C1}{$C_1/C_s$}
\includegraphics[width=4in]{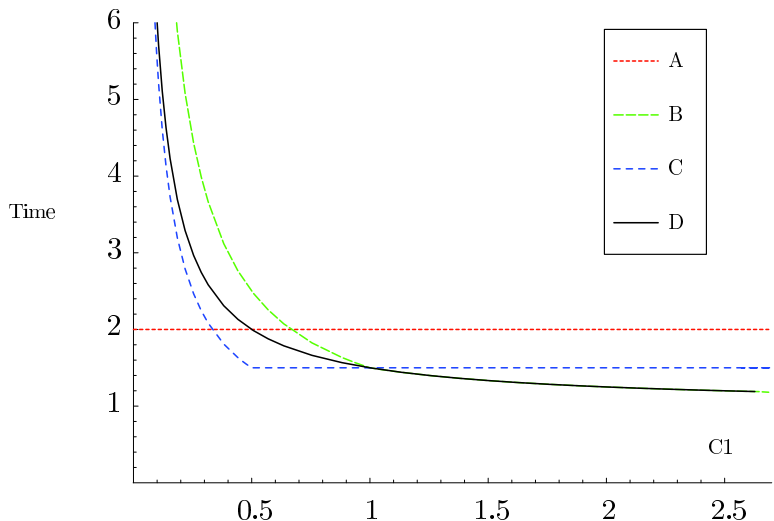}
\caption{Minimal makespan as a function of $C_1/C_S$ in the four possible cases when $N=M=2$.}\label{f:graph1}
\end{psfrags}
\end{center}
\end{figure}

The optimal strategy arises from A, C or D as $C_1/C_S$ lies in the intervals
$[0,1/3]$, $[1/3,1]$ or $[1,\infty)$ respectively. In $[1,\infty)$, B and D
yield the same. See Figure~\ref{f:graph1}. Note that under the optimal schedule
for case C one peer has to wait while the other starts downloading. This
illustrates that greedy-type distributed algorithms may not be optimal and that
restricting uploaders to a single upload is sometimes necessary for an optimal
scheduling (cf.~Section~\ref{s:efdmodel}).

\subsection{Insight for special cases with large $M$}\label{ss:efdcentrfluidgenspecial}

We still assume $C_1=C_2=\cdots=C_N$, but $C_S$ might be different. In the
limiting case that the file can be divided into infinitely many parts, the
problem can be easily solved for any number $N$ of users. Let each user
download a fraction $1-\alpha$ directly from the server at rate ${C_S}/{N}$ and
a fraction $\alpha/(N-1)$ from each of the other $N-1$ peers, at rate
$\min\{{C_S}/{N},{C_1}/(N-1)\}$ from each. The makespan is minimized by
choosing $\alpha$ such that the times for these two downloads are equal, if
possible. Equating them, we find the minimal makespan as follows.\medskip

\begin{figure}[h]
\begin{psfrags}
\psfrag{Time}{$\dfrac{\text{Time}}{C_s}$} \psfrag{C1}{$C_1/C_s$}
\psfrag{M=1}{$M=1$} \psfrag{M=2}{$M=2$} \psfrag{M=3}{$M=\infty$}
\centering
\includegraphics[width=4in]{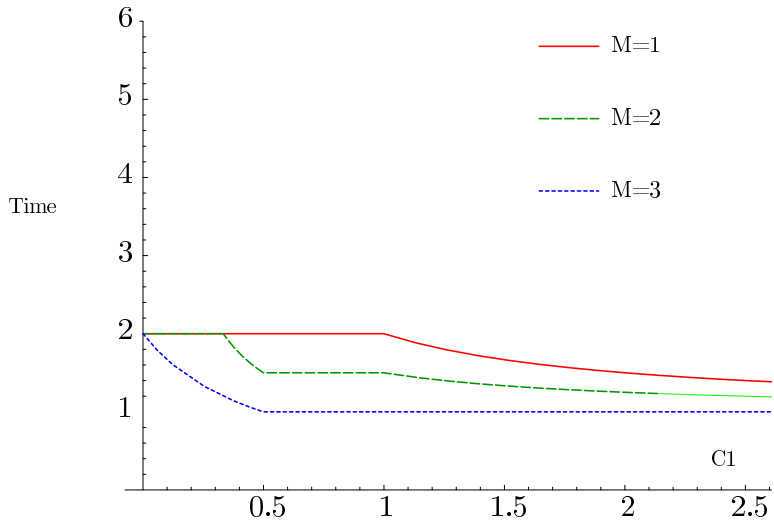}
\caption{Minimal makespan as a function of $C_1/C_S$ for different values of
$M$ when $N=2$.}\label{f:graph2}
\end{psfrags}
\end{figure}

\noindent {\bf Case 1:} $C_1/(N-1) \leq C_S/{N}$:
\begin{equation}
\frac{(1-\alpha)N}{C_S}=\frac{\alpha}{C_1} \quad\Longrightarrow\quad
\alpha=\frac{NC_1}{C_S+NC_1} \quad\Longrightarrow\quad
T=\frac{N}{C_S+NC_1}\,.
\end{equation}
\noindent {\bf Case 2:} $C_1/(N-1) \geq C_S/{N}$:
\begin{equation}
\frac{(1-\alpha)N}{C_S}=\frac{\alpha N}{(N-1)C_S}
\quad\Longrightarrow\quad \alpha=\frac{N-1}{N}
\quad\Longrightarrow\quad T=\frac{1}{C_S}\,.
\end{equation}
In total, there are $N$ MB to upload and the total available upload capacity is
$C_S+NC_1$ MBps. Thus, a lower bound on the makespan is ${N}/(C_S+NC_1)$
seconds. Moreover, the server has to upload his file to at least one user.
Hence another lower bound on the makespan is $1/C_S$. The former bound
dominates in case 1 and we have shown that it can be achieved. The latter bound
dominates in case 2 and we have shown that it can be achieved. As a result, the
minimal makespan is
\begin{equation}\label{e:efdfluidsimple}
T^*= \max \left\{ \frac{1}{C_S},\frac{N}{C_S+NC_1} \right\}\,.
\end{equation}
Figure~\ref{f:graph2} shows the minimal makespan when the file is split in 1, 2
and infinitely many file parts when $N=2$. It illustrates how the makespan
decreases with $M$.

In the next section, we extend the results in this limiting case to a much more
general scenario.

\section{Centralized Fluid Limit Solution}\label{s:efdcentrfluidgen}

In this section, we generalize the results of
Section~\ref{ss:efdcentrfluidgenspecial} to allow for general capacities $C_i$.
Moreover, instead of limiting the number of sources to one designated server
with a file to disseminate, we now allow every user $i$ to have a file that is
to be disseminated to all other users. We provide the centralized solution in
the limiting case that the file can be divided into infinitely many parts.

Let $F_i\geq 0$ denote the size of the file that user $i$ disseminates to all
other users. Seeing that in this situation there is no longer one particular
server and everything is symmetric, we change notation for the rest of this
section so that there are $N\geq 2$ users $1, 2, \ldots, N$. Moreover, let $F=
\sum_{i=1}^N F_i$ and $C= \sum_{i=1}^N C_i$. We will prove the following
result.

\begin{theorem}
In the fluid limit, the minimal makespan is
\begin{equation}\label{e:genfluidsol}
T^* = \max \left\{ \frac{F_1}{C_1}, \frac{F_2}{C_2}, \ldots , \frac{F_N}{C_N}, \frac{(N-1)F}{C} \right\}
\end{equation}
and this can be achieved with a two-hop strategy, i.e., one in which users
$i$'s file is uploaded to user $j$, either directly from user $i$, or via at
most one intermediate user.
\end{theorem}
\begin{proof}
\noindent The result is obvious for $N=2$. Then the minimal makespan is $\max
\{ F_1/C_1, F_2/C_2 \}$ and this is exactly the value of $T^*$ in
\eqref{e:genfluidsol}.\medskip

So we consider $N\geq3$. It is easy to see that each of the $N+1$ terms within
the braces on the right hand side of \eqref{e:genfluidsol} are lower bounds on
the makespan. Each user has to upload his file at least to one user, which
takes time $F_i/C_i$. Moreover, the total volume of files to be uploaded is
$(N-1)F$ and the total available capacity is $C$. Thus, the makespan is at
least $T^*$, and it remains to be shown that a makespan of $T^*$ can be
achieved. There are two cases to consider.\medskip

\noindent {\bf Case 1:} $(N-1)F/C \geq \max_i F_i/C_i$ for all $i$.

\noindent In this case, $T^*=(N-1)F/C$. Let us consider the 2-hop strategy in
which each user uploads a fraction $\alpha_{ii}$ of its file $F_i$ directly to
all $(N-1)$ peers, simultaneously and at equal rates. Moreover, he uploads a
fraction $\alpha_{ij}$ to peer $j$ who in turn then uploads it to the remaining
$(N-2)$ peers, again simultaneously and at equal rates. Note that $\sum_{j=1}^N
\alpha_{ij}=1$.\medskip

\noindent Explicitly constructing a suitable set
$\alpha_{ij}$, we thus obtain the problem
\begin{equation}
\min T
\end{equation}
subject to, for all $i$,
\begin{equation}
\frac{1}{C_i} \left[ \alpha_{ii}F_i(N-1) + \sum_{k\neq i} \alpha_{ik} F_i +
\sum_{k\neq i} \alpha_{ki} F_k (N-2) \right] \leq T\,.
\end{equation}
We minimize $T$ by choosing the $\alpha_{ij}$ in such a way as to equate the
$N$ left hand sides of the constraints, if possible. Rewriting the expression
in square brackets, equating the constraints for $i$ and $j$ and then summing
over all $j$ we obtain
\begin{align}
&C \left[ \alpha_{ii}F_i(N-2) + F_i + \sum_{k\neq i} \alpha_{ki} F_k (N-2) \right]\nonumber\\
&= C_i \left[ (N-2)\sum_j \alpha_{jj}F_j + F + (N-2)(F-\sum_j \alpha_{jj}F_j) \right]\nonumber\\
&= (N-1)C_iF.
\end{align}
Thus,
\begin{eqnarray}
\alpha_{ii}F_i(N-2) + F_i + \sum_{k\neq i} \alpha_{ki} F_k (N-2)= (N-1)\frac{C_i}{C}F.
\end{eqnarray}
Note that there is a lot of freedom in the choice of the $\alpha$ so let us
specify that we require $\alpha_{ki}$ to be constant in $k$ for $k\neq i$, that
is $\alpha_{ki}=\alpha^*_{i}$ for $k\neq i$. This means that $i$ has the
capacity to take over a certain part of the dissemination from some peer, then
it can and will also take over the same proportion from any other peer. Put
another way, user $i$ splits excess capacity equally between its peers. Thus,
\begin{eqnarray}
\alpha_{ii}F_i(N-2) + F_i + \alpha^*_{i} (N-2) (F-F_i) = (N-1)\frac{C_i}{C}F
\end{eqnarray}
Still, we have twice as many variables as constraints. Let us also specify that
$\alpha^*_{i}=\alpha_{ii}$ for all $i$. Similarly as above, this says that the
proportion of its own file $F_i$ that $i$ uploads to all its peers (rather than
just to one of them) is the same as the proportion of the files that it takes
over from its peers. Then
\begin{eqnarray}
\alpha^*_{i}=\frac{(N-1)(C_i/C)F - F_i}{(N-2)F} = \frac{(N-1)C_i}{(N-2)C}-\frac{F_i}{(N-2)F},
\end{eqnarray}
where $\sum_i \alpha^*_{i}=1 $ and $\alpha^*_{i} \geq 0$, because in case 1
$F_i/C_i \leq (N-1)F/C$. \medskip

With these $\alpha_{ij}$, we obtain the time for $i$ to complete its upload and
hence the time for everyone to complete their upload as
\begin{eqnarray}
T&=& \frac{1}{C_i} \left[ \alpha^*_{i}F_i(N-2) + F_i + \sum_{k\neq i} \alpha^*_{i} F_k (N-2) \right]\nonumber\\
&=& \frac{(N-1)F_i}{C} - \frac{{F_i}^2}{C_iF} +\frac{F_i}{C_i} + \frac{(N-1)(F-F_i)}{C} - \frac{F_i(F-F_i)}{C_iF}\nonumber\\
&=& (N-1)F/C.
\end{eqnarray}
Note that there is no problem with precedence constraints. All uploads happen
simultaneously stretched out from time $0$ to $T$. User $i$ uploads to $j$ a
fraction $\alpha_{ij}$ of $F_i$. Thus, he does so at constant rate
$\alpha_{ij}F_i/T_i=\alpha_{ij}F_i/T$. User $j$ passes on the same amount of
data to each of the other users in the same time, hence at the same rate
$\alpha_{ij}F_i/T_j=\alpha_{ij}F_i/T$.\medskip

Thus, we have shown that if the aggregate lower bound dominates the others, it
can be achieved. It remains to be shown that if an individual lower bound
dominates, than this can be achieved also.\medskip

\noindent {\bf Case 2:} $F_i/C_i > (N-1)F/C$ for some $i$.

\noindent By contradiction it is easily seen that this cannot be the case for
all $i$. Let us order the users in decreasing order of $F_i/C_i$, so that
$F_1/C_1$ is the largest of the $F_i/C_i$. We wish to show that all files can
be disseminated within a time of $F_1/C_1$. To do this we construct new
capacities $C'_i$ with the following properties:
\begin{gather}
C'_1=C_1,\label{epropa}\\
C'_i \leq C_i \text{ for } i \neq 1,\label{epropb}\\
(N-1)F/C'=F_1/C'_1=F_1/C_1 \text{ and}\label{epropc}\\
F_i/C_i' \leq F_1/C_1.\label{epropd}
\end{gather}
This new problem satisfies the condition of Case 1 and so the minimal makespan
is $T'=F_1/C_1$. Hence the minimal makespan in the original problem is
$T=F_1/C_1$ also, because the unprimed capacities are greater or equal to the
primed capacities by property \eqref{epropb}.\medskip

To explicitly construct capacities satisfying
\eqref{epropa}--\eqref{epropd}, let us define
\begin{equation}
C'_i= (N-1) \frac{C_1}{F_1} \gamma_i F_i
\end{equation}
with constants $\gamma_i \geq 0$ such that
\begin{equation}\label{e:gamma1}
\sum_i \gamma_i F_i = F\,.
\end{equation}
Then $(N-1)F/C'=F_1/C_1$, that is \eqref{epropc} holds. Moreover, choosing
\begin{equation}\label{e:gamma2}
\gamma_i \leq \frac{1}{N-1}\frac{C_i}{F_i}\frac{F_1}{C_1}
\end{equation}
ensures $C'_i \leq C_i$, i.e.~property \eqref{epropb} and choosing
\begin{equation}\label{e:gamma3}
\gamma_i \geq \frac{1}{N-1}
\end{equation}
ensures $F_i/C'i \leq F_1/C_1$, that is property \eqref{epropd}. Furthermore,
the previous two conditions together ensure that $\gamma_1=1/(N-1)$ and thus
$C'_1=C_1$, that is property \eqref{epropa}. It remains to construct a set of
parameters $\gamma_i$ that satisfies \eqref{e:gamma1}, \eqref{e:gamma2} and
\eqref{e:gamma3}.\medskip\\
Putting all $\gamma_i$ equal to the lower bound \eqref{e:gamma3} gives $\sum_i
\gamma_i F_i = F/(N-1) $, that is too small to satisfy \eqref{e:gamma1}.
Putting all equal to the upper bound \eqref{e:gamma2} gives $ \sum_i \gamma_i
F_i = F_1C/(N-1)C_1 $, that is too large to satisfy \eqref{e:gamma1}. So we
pick a suitably weighted average instead. Namely,
\begin{equation}
\gamma_i = \frac{1}{N-1} \left[ \delta \frac{C_i}{F_i}\frac{F_1}{C_1} + (1-\delta) \right]
\end{equation}
such that
\begin{equation}
\delta \frac{C}{N-1}\frac{F_1}{C_1} + (1-\delta)\frac{F}{N-1}=F
\end{equation}
that is
\begin{equation}
\delta = \frac{(N-2)FC_1}{F_1C-FC_1}\,.
\end{equation}
Substituting back in we obtain
\begin{equation}
\gamma_i = \frac{1}{N-1} \frac{(N-2)FF_1C_i+F_iF_1C-(N-1)FF_iC_1}{(F_1C-FC_1)F_i}
\end{equation}
and thus
\begin{equation}
C'_i= \frac{C_1}{F_1} \frac{(N-2)FF_1C_i+F_iF_1C-(N-1)FF_iC_1}{F_1C-FC_1}\,
\end{equation}

By construction, these $C'_i$ satisfy properties \eqref{epropa}--\eqref{epropd}
and hence, by the results in Case 1, $T'=F_1/C_1$. Hence the minimal makespan
in the original problem $T=F_1/C_1$ also.
\end{proof}
\medskip

It is worth noting that there is a lot of freedom in the choice of the
$\alpha_{ij}$. We have chosen a symmetric approach, but other choices are
possible.\medskip

In practice, the file will not be infinitely divisible. However, we often have
$M >> \log(N)$ and this appears to be sufficient for \eqref{e:genfluidsol} to
be a good approximation. Thus, the fluid limit approach of this section is
suitable for typical and for large values of $M$.

\section{Decentralized Solution for Equal Capacities}\label{ss:efddecentreq}

In order to give a lower bound on the minimal makespan, we have been assuming a
centralized controller does the scheduling. We now consider a naive randomized
strategy and investigate the loss in performance that is due to the lack of
centralized control. We do this for equal capacities and in two different
information scenarios, evaluating its performance by analytic bounds,
simulation as well as direct computation. In Section~\ref{ss:efddecentreqone}
we consider the special case of one file part, in Section~\ref{ss:efddecentgen}
we consider the general case of $M$ file parts. We find that even this naive
strategy disseminates the file in an expected time whose growth rate with $N$
is similar to the growth rate of the minimal time that we have found for a
centralized controller (cf. Section \ref{s:efdcentreq}). This suggests that the
performance of necessarily decentralized P2P file dissemination systems should
still be close to our performance bounds so that they are useful in practice.

\subsection{The special case of one file part}\label{ss:efddecentreqone}

\subsubsection*{Assumptions}

Let us start with the case $M=1$. We must first specify what information is
available to users. It makes sense to assume that each peer knows the number of
parts into which the file is divided, $M$, and the address of the server.
However, a peer might not know $N$, the total number of peers, nor its peers'
addresses, nor if they have the file, nor whether they are at present occupied
uploading to someone else.\medskip

We consider two different information scenarios. In the first one,
\emph{List}, the number of peers holding the file and their addresses
are known. In the second one, \emph{NoList}, the number and addresses
of all peers are known, but not which of them currently hold the
file. Thus, in \emph{List}, downloading users choose uniformly at
random between the server and the peers already having the file. In
\emph{NoList}, downloading users choose uniformly amongst the server
and all their peers. If a peer receives a query from a single peer, he
uploads the file to that peer. If a peer receives queries from
multiple peers, he chooses one of them uniformly at random. The others
remain unsuccessful in that round. Thus, in \emph{List} transmission
can fail only if too many users try to download simultaneously from
the same uploader. In \emph{NoList}, transmission might also fail if a
user tries to download from a peer who does not yet have the file.

\subsubsection*{Theoretical Bounds}

The following theorem explains how the expected makespan that is
achieved by the randomized strategy grows with $N$, in both the
\emph{List} and the \emph{NoList} scenarios.

\begin{theorem}\label{thm:decbound}
In the uplink-sharing model, with equal upload capacities, the
expected number of rounds required to disseminate a single file to all
peers in either the {List} or {NoList} scenario is $\Theta(\log N)$.
\end{theorem}

\begin{proof}
In the \emph{List} scenario our simple randomized algorithm runs in
less time than in the \emph{NoList} scenario. Since already have the
lower bound given by Theorem~\ref{thm:makespan}, it suffices to prove
that the expected runing time in the \emph{NoList} scenario is $O(\log
N)$. There is also similar direct proof that the expected running time
under the \emph{List} scenario is $O(\log N)$.

Suppose we have reached a stage in the dissemination at which $n_1$
peers (including the server) have the file and $n_0$ peers do not,
with $n_0+n_1=N+1$. (The base case is $n_1=1$, when only the server
has the file.)  Each of the peers that does not have the file randomly
chooses amongst the server and all his peers (\emph{NoList}) and tries
to download the file. If more than one peer tries to download from the
same place then only one of the downloads is successful. The proof has
two steps.

(i) Suppose that $n_1\leq n_0$. Let $i$ be the server or a peer who has the
file and let $I_i$ be an indicator random variable that is 0 or 1 as $i$ does
or does not upload it.  Let $Y=\sum_i I_i$, where the sum is taken over all
$n_1$ peers who have the file. Thus $n_1-Y$ is the number of uploads that take
place. Then
\begin{equation}
EI_i=\left(1-\frac{1}{N}\right)^{n_0}\leq
\left(1-\frac{1}{2n_0}\right)^{n_0}\leq \frac{1}{\sqrt{e}}\,.
\end{equation}
Now since $E(\sum_i I_i)=\sum_i EI_i$,
we have $EY\leq n_1/\sqrt{e}$.
Thus, by the Markov inequality, that for a nonnegative random variable $Y$ we
have that for any $k$ (not necessarily an integer) $P(Y\geq k)\leq (1/k)EY$,
we have by taking $k= (2/3)n_1$,
\begin{equation}
P\bigl(\text{$n_1-Y\equiv$ number of uploads $\leq
\tfrac{1}{3}n_1$}\bigr)=P(Y\geq \tfrac{2}{3}n_1) \leq
\frac{n_1/\sqrt{e}}{\tfrac{2}{3}n_1} = 3/(2\sqrt{e}) <1\,.\label{oo1}
\end{equation}
Thus the expected number of steps required for the number of peers who
have the file to increases from $n_1$ to at least $n_1+ (1/3)n_1 =
(4/3)n_1$ is bounded by a geometric random variable with mean
$\mu=1/(1-3/(2\sqrt{e}))$. This implies that we will reach a state in
which more peers have the file than do not in an expected time that is
$O(\log N)$. From that point we continue with step (ii) of the proof.

(ii) Suppose $n_1>n_0$.  Let $j$ be a peer who does not have the file and let
$J_j$ be an indicator random variable that is 0 or 1 as peer $j$ does or does
not succeed in downloading it.  Let $Z=\sum_j J_j$, where the sum is taken over
all $n_0$ peers who do not have the file. Suppose $X$ is the number of the
other $n_0-1$ peers that try to download from the same place as does peer $j$.
Then
\begin{align}
P(J_j=0)&=E\left[\frac{n_1}{N}\left(\frac{1}{1+X}\right)\right]\nonumber\\
&\geq E\left[\frac{n_1}{N}\left(1-X\right)\right]\nonumber\\
&=\frac{n_1}{N}\left(1-\frac{n_0-1}{N}\right)\nonumber\\
&=\frac{n_1}{N}\left(1-\frac{N-n_1}{N}\right)\nonumber\\
&=\frac{n_1^2}{N^2}\nonumber\\
&\geq 1/4\,.
\end{align}
Hence $EZ\leq (3/4)n_0$ and so, again using the Markov inequality,
\begin{equation}
P\bigl(\text{$n_0-Z\equiv$ number of downloads $\leq \tfrac{1}{8}n_0$}\bigr)=
P\bigl(Z\geq\tfrac{7}{8}n_0\bigr)\leq
\frac{\tfrac{3}{4}n_0}{\tfrac{7}{8}n_0}=\tfrac{6}{7}\,.
\end{equation} It
follows that the number of peers who do not yet have the file
decreases from $n_0$ to no more than $(7/8)n_0$ in an expected number
of steps no more than $\mu'=1/(1-\tfrac{6}{7})=7$. Thus the number of
steps needed for the number of peers without the file to decrease from
$n_0$ to $0$ is $O(\log n_0)=O(\log N)$.  In fact, this is a weak
upper bound. By more complicated arguments we can show that if
$n_0=aN$, where $a\leq 1/2$, then the expected remaining time for our
algorithm to complete under \emph{NoList} is $\Theta(\log\log N)$.
For $a>1/2$ the expected time remains $\Theta(\log N)$.
\end{proof}

\subsubsection*{Simulation}

For the problem with one server and $N$ users we have carried out $1000$
independent simulation runs\footnote{As many as 1000 runs were required for the
comparison with the computational results in Tables~\ref{t:list} and
\ref{t:nolist}, mainly because the makespan always takes integer values.} for a
large range of parameters, $N=2$, $4$, \ldots, $2^{25}$. We found that the
achieved expected makespan appears to grow as $a + b \times \log_2{N}$.
Motivated by this and the theoretical bound from Theorem \ref{thm:decbound} we
fitted the linear model
\begin{equation}
y_{ij}=\alpha + \beta x_i + \epsilon_{ij}\,,\label{e:lm}
\end{equation}
where $y_{ij}$ is the makespan for $x_i=\log_2 2^i$, obtained in run $j$,
$j=1,\ldots,1000$.
Indeed, the model fits the data very well in both scenarios. We obtain the
following results that enable us to compare the expected makespan of the naive
randomized strategy to the that of a centralized controller.

For \emph{List}, the regression analysis gives a good fit, with Multiple
R-squared value of $0.9975$ and significant p- and t-values.
The makespan increases as
\begin{equation}
1.1392 + 1.1021 \times \log_2{N}\,. \label{Lreg}
\end{equation}
For \emph{NoList}, there is more variation in the data than for \emph{List},
but, again, the linear regression gives a good fit, with Multiple R-squared of
$0.9864$ and significant p- and t-values.
The makespan increases as
\begin{equation}
1.7561 + 1.5755 \times \log_2{N}\,.\label{NLreg}
\end{equation}
As expected, the additional information for \emph{List} leads to a
significantly lesser makespan when compared to \emph{NoList}, in particular the
log-term coefficient is significantly smaller. In the \emph{List} scenario, the
randomized strategy achieves a makespan that is very close to the centralized
optimum of $1+\lfloor\log_2{N}\rfloor$ of Section \ref{s:efdcentreq}: It is
only suboptimal by about 10\%. Hence even this simple randomized strategy
performs well in both cases and very well when state information is available,
suggesting that our bounds are useful in practice.

\subsubsection*{Computations}

Alternatively, it is possible to compute the mean makespan analytically by
considering a Markov Chain on the state space ${0,1,2,\ldots,N}$, where state
$i$ corresponds to $i$ of the $N$ peers having the file. We can calculate the
transition probabilities $p_{ij}$. In the \emph{NoList} case, for example,
following the Occupancy Distribution (e.g., \cite{b:jkk93}), we obtain
\begin{equation}
p_{ii+m} = \sum_{j=i-m}^i \frac{(-1)^{j-i+m}i!}{(i-j)!(i-m)!(j-i+m)!}
\left(\frac{N-1-j}{N-1}\right)^{N-i}\,.
\end{equation}
Hence we can successively compute the expected hitting times $k(i)$ of state
$N$ starting from state $i$ via
\begin{equation}
k(i) = \frac{1+ \sum_{j>i}k(j)p_{ij}}{1-p_{ii}}\,.
\end{equation}
The resulting formula is rather complicated, but can be evaluated exactly using
arbitrary precision arithmetic on a computer. Computation times are long, so to
keep them shorter we only work out the transition probabilities of the
associated Markov Chain exactly. Hitting times are then computed in double
arithmetic, that is, to 16 significant digits.
Even so, computations are only feasible up to $N=512$ with our equipment,
despite repeatedly enhanced efficiency. This suggests that simulation is the
more computationally efficient approach to our problem. The computed mean
values for \emph{List} and \emph{NoList} are shown in Tables~\ref{t:list} and
\ref{t:nolist} respectively. The difference to the simulated values is small
without any apparent trend.
It can also be checked by computing the standard deviation that the computed
mean makespan is contained in the approximate 95\% confidence interval of the
simulated mean makespan. The only exception is for $N=128$ for \emph{NoList}
where it is just outside by approximately $0.0016$.\medskip

Thus, the computations prove our simulation results accurate. Since simulation
results are also obtained more efficiently, we shall stick to simulation when
investigating the general case of $M$ file parts in the next section.

\begin{table}[h]\small
\renewcommand{\tablename}{\small Table\centering}
\begin{minipage}{.45\textwidth}
\[
\begin{array}{rrrr}
\hline N & \text{sim.} & \text{comp.} & \text{difference}\\
\hline 2 & 2.000 & 2.000 & =0.000\\ 4 & 3.089 & 3.083 & +0.006\\ 8 &
4.167 & 4.172 & -0.005\\ 16 & 5.333 & 5.319& +0.014\\ 32 & 6.534 &
6.538 & -0.004\\ 64 & 7.806 & 7.794 & +0.012\\ 128 & 8.994 & 8.981 &
+0.013\\ 256 & 10.059 & 10.057 & +0.002\\ 512 & 11.107 & 11.116 &
-0.009\\ \hline
\end{array}
\]
\caption{\small Simulated and computed mean makespans for
\emph{List} are close.}\label{t:list}
\end{minipage}\hfill
\begin{minipage}{.45\textwidth}
\[
\begin{array}{rrrr}
\hline N & \text{sim.} & \text{comp.} & \text{difference}\\
\hline 2 & 2.314 & 2.333 & -0.019\\ 4 & 4.071 & 4.058 & +0.013\\ 8 &
5.933 & 5.956 & -0.023\\ 16 & 7.847 & 7.867 & -0.020\\ 32 & 9.689 &
9.710 & -0.021\\ 64 & 11.430 & 11.475 &-0.045\\ 128& 13.092 & 13.173 &
-0.081\\ 256& 14.827 & 14.819 &+0.008\\ 512& 16.426 & 16.427 &
-0.001\\ \hline
\end{array}
\]
\caption{\small Simulated and computed mean makespans for
\emph{NoList} are close.}\label{t:nolist}
\end{minipage}
\end{table}

\subsection{The general case of $M$ file parts}\label{ss:efddecentgen}

\subsubsection*{Assumptions}

We now consider splitting the file into several file parts. With the same
assumptions as in the previous section, we repeat the analysis for \emph{List}
for various values of $M$. Thus, in each round, a downloading user connects to
a peer chosen uniformly at random from those peers that have at least one file
part that the user does not yet have. An uploading peer randomly chooses one
out of the peers requesting a download from him. He uploads to that peer a file
part that is randomly chosen from amongst those that he has and the peer still
needs.

\subsubsection*{Simulation}

Again, we consider a large range of parameter. We carried out 100 independent
runs for each $N=2$, $4$, \ldots, $2^{15}$. For each value of $M=1-5$, $8$,
$10$,  $15$, $20$, $50$ we fitted the linear model \eqref{e:lm}.

Table~\ref{t:listsimM} summarizes the simulation results. The Multiple
R-squared values indicate a good fit, although the fact that these decrease
with $M$ suggests there may be a finer dependence on $M$ or $N$. In fact, we
obtain a better fit using Generalized Additive Models (cf.~\cite{b:ht90}).
However, our interest here is not in fitting the best possible model, but to
compare the growth rate with $N$ to the one obtained in the centralized case in
Section~\ref{s:efdcentreq}. Moreover, from the diagnostic plots
we note that the actual performance for large $N$ is better than given by the
regression line, increasingly so for increasing $M$. In each case, we obtain
significant p- and t-values. The regression $0.7856 + 1.1520 \times \log_2{N}$
for $M=1$ does not quite agree  with $1.1392 + 1.1021 \times \log_2{N}$ found
in \eqref{Lreg}. It can be checked, by repeating the analysis there for $N=2$,
$4$, \ldots, $2^{15}$ that this is due to the different range of $N$. Thus, our
earlier  result of $1.1021$ might be regarded more reliable, being based on $N$
ranging up to $2^{25}$.\medskip

\begin{table}[h]\small
\[
\begin{array}{rlcc}
\hline M & \multicolumn{1}{c}{\text{Fitted}} & \text{Multiple}&
1/M\\ &\multicolumn{1}{c}{\text{makespan}} &
\multicolumn{1}{c}{\text{R-squared}}\\ \hline 1 & 0.7856 + 1.1520
\times \log_2{N} & 0.9947 &1.000\\ 2 & 1.3337 + 0.6342 \times
\log_2{N} & 0.9847 &0.500\\ 3 & 1.4492 + 0.4561 \times \log_2{N} &
0.9719 &0.333\\ 4 & 1.4514 + 0.3661 \times \log_2{N} & 0.9676 &0.250\\
5 & 1.4812 + 0.3045 \times \log_2{N} & 0.9690 &0.200\\ 8 & 1.4907 +
0.2113 \times \log_2{N} & 0.9628 &0.125\\ 10& 1.4835 + 0.1791 \times
\log_2{N} & 0.9602 &0.100\\ 15& 1.4779 + 0.1326 \times \log_2{N} &
0.9530 &0.067\\ 20& 1.4889 + 0.1062 \times \log_2{N} & 0.9449 &0.050\\
50& 1.4524 + 0.0608 \times \log_2{N} & 0.8913 &0.020\\ \hline
\end{array}
\]
\caption{Simulation results in the decentralized \emph{List} scenario for
various values of $M$ and log-term coefficients in the centralized optimum
(cf.~Theorem \ref{thm:makespan}).\label{t:listsimM}}
\end{table}

\begin{figure}[h]
\begin{psfrags}
\psfrag{M}{\small $M$}
\psfrag{makespan coefficient}{\small makespan coefficient}
\psfrag{Log-term coefficient}{Log-term coefficient}
\centering \includegraphics[width=4in]{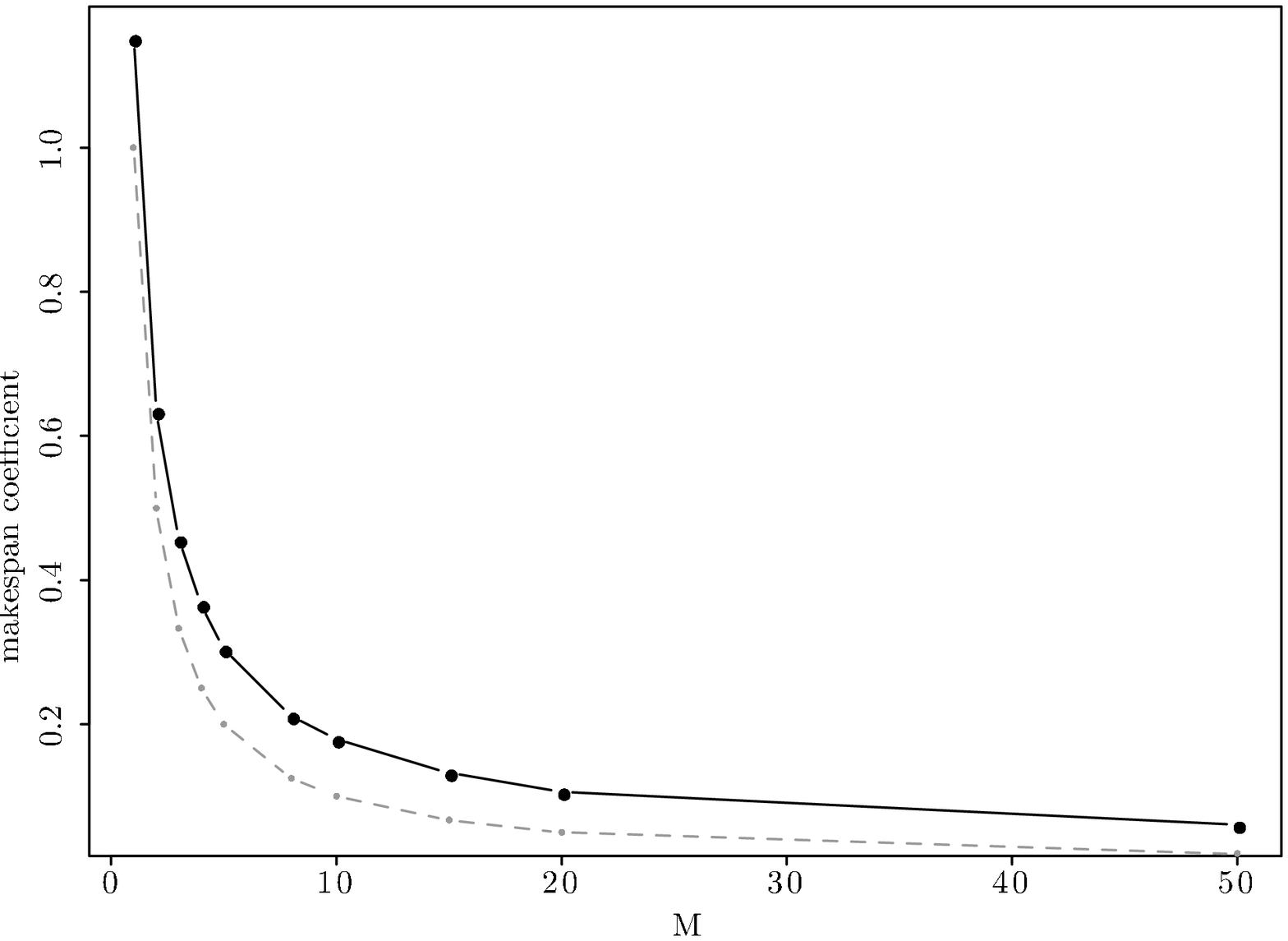}
\caption{Illustration of the log-term coefficients of the makespan from
Table~\ref{t:listsimM}: the decentralized \emph{List} scenario (solid) and the
idealized centralized scenario (dashed).} \label{f:gaincomparison}
\end{psfrags}
\end{figure}

We conclude that, as in the centralized scenario, the makespan can also be
reduced significantly in a decentralized scenario even when a simple randomized
strategy is used to disseminate the file parts. However, as we note by
comparing the second and fourth columns of Table~\ref{t:listsimM}, as $M$
increases the achieved makespan compares less well relative to the centralized
minimum of $1+(1/M)\lfloor\log_2 N\rfloor$. In particular, note the slower
decrease of the log-term coefficient. This is depicted in
Figure~\ref{f:gaincomparison}.

Still, we have seen that even this naive randomized strategy disseminates the
file in an expected time whose growth rate with $N$ is similar to the growth
rate of the minimal time that we have found for a centralized controller in
Section \ref{s:efdcentreq}, confirming our performance bounds are useful in
practice. This is confirmed also by initial results of current work on the
performance evaluation of the Bullet' system \cite{krav03}.

The program code for simulations as well as the computations and the diagnostic
plots used in this section are available on request and will be made available
via the Internet\footnote{http://www.statslab.cam.ac.uk/$\sim$jm288/}.

\section{Discussion}\label{s:efddisc}

In this paper, we have given three complementary solutions for the minimal time
to fully disseminate a file of $M$ parts from a server to $N$ end users in a
centralized scenario, thereby providing a lower bound on and a performance
benchmark for P2P file dissemination systems.
Our results illustrate how the P2P approach, together with splitting the file
into $M$ parts, can achieve a significant reduction in makespan. Moreover, the
server has a reduced workload when compared to the traditional client/server
approach in which it does all the uploads itself.
We also investigate the part of the loss in efficiency that is due to the lack
of centralized control in practice. This suggests that the performance of
necessarily decentralized P2P file dissemination systems should still be close
to our performance bound confirming their practical use.

It would now be very interesting to compare dissemination times of the various
efficient real overlay networks directly to our performance bound. A
mathematical analysis of the protocols is rarely tractable, but simulation or
measurements such as in~\cite{iubfag04} and~\cite{pges05} for the BitTorrent
protocol can be carried out in an environment suitable for this comparison. Cf.
also testbed results for Slurpie \cite{sbb04} and simulation results for
Avalanche \cite{gr05}. It is current work to compare our bounds to the makespan
obtained by Bullet'~\cite{krav03}. Initial results confirm their practical use
further.

In practice, splitting the file and passing on extra information has an
overhead cost. Moreover, with the Transmission Control Protocol (TCP), longer
connections are more efficient than shorter ones. TCP is used practically
everywhere except for the Internet Control Message Protocol (ICMP) and User
Datagram Protocol (UDP) for real-time applications. For further details see
\cite{b:s04}.
Still, with an overhead cost it will not be optimal to increase $M$ beyond a
certain value. This could be investigated in more detail.

In the proof of Lemma 1 and Lemma 2 we have used fair sharing and continuity
assumptions. It would be of interest to investigate whether one of them or both
can be relaxed.

It would be interesting to generalize our results to account for a dynamic
setting with peers arriving and perhaps leaving when they have completed the
download of the file. In Internet applications users often connect for only
relatively short times. Work in this direction, using a fluid model to study
the steady-state performance, is pursued in \cite{qs04} and there is other
relevant work in \cite{yd05}.

Also of interest would be to extend our model to consider users who prefer to
free-ride and do not wish to contribute uploading effort. Or, to users who
might want to leave the system once they have downloaded the whole file, a
behaviour sometimes referred to as \emph{easy-riding}. The BitTorrent protocol,
for example, implements a choking algorithm to limit free-riding.

In another scenario it might be appropriate to assume that users \emph{push}
messages rather than \emph{pull} them. See \cite{fz97} for an investigation of
the design space for distributed information systems. The push-pull distinction
is also part of their classification. In a push system, the centralized case
would remain the same. However, we expect the decentralized case to be
different. There are a number of other interesting questions which could be
investigated in this context. For example, what happens if only  a subset of
the users is actually interested in the file, but the uploaders do not know
which.

From a mathematical point of view it would also be interesting to consider
additional download constraints explicitly as part of the model, in particular
when up- and download capacities are all different and not positively
correlated. We might suppose that user $i$ can upload at a rate $C_i$ and
simultaneously download at rate $D_i$.

More generally, one might want to assume different capacities for all links
between pairs. Or, phrased in terms of transmission times, let us assume that
for a file to be sent from user $i$ to user $j$ it takes time $t_{ij}$.
Then we obtain a transportation network, where instead of link costs we now
have link delays. This problem can be phrased as a one-to-all shortest path
problem if $C_j$ is at least $N+1$. This suggests that there might be some
relation which could be exploited. On the other hand, the problem is
sufficiently different so that greedy algorithms, induction on nodes and
Dynamic Programming do not appear to work. Background on these can be found in
\cite{b:bt97} and \cite{b:bhs05}. For $M=1$, Pr\"{u}fer's $(N+1)^{N-1}$
labelled trees~\cite{b:b98} together with the obvious $O(N)$ algorithm for the
optimal scheduling given a tree is an exhaustive search. A Branch and Bound
algorithm can be formulated.

\nocite{mw04}
\nocite{mw04dagStress}
\bibliographystyle{abbrv}
\bibliography{GenCapPerf05Full}
\end{document}